\documentclass[preprint,floatfix,amsmath,amssymb,aip,apl]{revtex4-1}
\usepackage{graphicx,float,amssymb,subfigure,epstopdf}
\usepackage{dcolumn}
 \usepackage{CJK} 
\begin{document}
 \title{Spin galvanic effect at the conducting SrTiO$_3$ surfaces}
 \author{Wenxu Zhang\footnote{Corresponding author. E-mail address: xwzhang@uestc.edu.cn}, Qiuru Wang, Bin Peng, Huizhong Zeng}
  \affiliation{State Key Laboratory of Electronic Thin Films and Integrated Devices,
  	University of Electronic Science and Technology of China, Chengdu, 610054, P. R. China}

 \author {Wee Tee Soh}
 \affiliation{Center for Superconducting and Magnetic Materials, Department of Physics, National University of Singapore, 2 Science Drive 3, Singapore 117551, Singapore} 
  \author{Chong Kim Ong}
 \affiliation{Center for Superconducting and Magnetic Materials, Department of Physics, National University of Singapore, 2 Science Drive 3, Singapore 117551, Singapore} 
  \affiliation{Temasek Laboratories, National University of Singapore, 5A Engineering Drive 2, Singapore 117411}
 \author{ Wanli Zhang}
 \affiliation{State Key Laboratory of Electronic Thin Films and Integrated Devices,
 University of Electronic Science and Technology of China, Chengdu, 610054, P. R. China}
 \date{\today}
\begin{abstract}
The (001) surface of SrTiO$_3$ were transformed from insulating to conducting after Ar$^+$ irradiation, producing a quasi two-dimensional electron gas (2DEG). This conducting surface layer can introduce Rashba spin orbital coupling due to the broken inversion symmetry normal to the plane. The spin splitting of such a surface has recently been demonstrated by magneto-resistance and angular resolved photoemission spectra measurements. Here we present experiments evidencing a large spin-charge conversion at the surface. We use spin pumping to inject a spin current from NiFe film into the surface, and measure the resulting charge current. The results indicate that the Rashba effect at the surface can be used for efficient charge-spin conversion, and the large efficiency is due to the multi-$d$-orbitals and surface corrugation. It holds great promise in oxide spintronics.   
\end{abstract}
\maketitle

It was pointed out by Dresselhaus\cite{dresselhaus} and Rashba\cite{rashba60}, in non-centrosymmetric zinc-blende or wurtzite semiconductors that the spin-orbital coupling (SOC) becomes odd in the electron momentum $\vec{p}$. The SO Hamiltonian can be written as $H_{SO}\sim \mu_B(\vec{E}\times\vec{p})\cdot\vec{\sigma}/mc^2$, where $\vec{E},\vec{p}, m, c, \vec{\sigma}$ and $\mu_B$ are the electric field, electron momentum, electron mass, light speed, Pauli matrices and Bohr magneton, respectively. As a consequence of this odd-in-$p$ SO coupling, electrons moving in an electric field experience an effective magnetic field $\vec{B}_{eff}\sim \vec{E}\times\vec{p}/mc^2$, which is coupled with the spin of electrons.
Due to the SO coupling, unpolarized charge currents can be converted into transverse pure spin currents, in a phenomenon called spin Hall effect (SHE). 
 Later on, it was pointed out by Bychkov and Rashba \cite{rashba84} in 1984 that if a crystal has a single high-symmetry axis and an invariant vector $\vec z$ oriented along this axis, similar SOC can be obtained. This is the so-called Rashba SOC (RSOC) with its associated electronic Hamiltonian written as $ H_{RSOC}=\alpha_R (\vec{\sigma}\times \vec{k})\cdot\vec{z}$, where $\vec{k}$ is the wave vector of the traveling electrons, and $\alpha_R$ is the so-called Rashba coupling constant. According to this, electrons moving in materials experience a wave vector dependent effective magnetic field with magnitude $B_{R}=2\alpha_R {k_F}/{g\mu_B}$ and its direction being dependent on $\vec k$ and $\vec z$, where $k_F$ is the wave vector at the Fermi level, $g$ is the electron Land\'{e} $g$-factor and $\mu_B=9.27\times 10 ^{-24}$J T$^{-1}$ is the Bohr magneton. Over the past 30 years, the RSOC has inspired a vast number of predictions, discoveries and innovative concepts. The Rashba physics in condensed matter was recently reviewed by Manchon et al\cite{manchon}. It was firstly experimentally demonstrated by Sih et al.\cite{sih} in a two-dimensional electron gas (2DEG) confined in (110) AlGaAs quantum wells using Kerr rotation microscopy. The Rashba coupling constant was determined to be $\alpha_R=1.8\times 10^{-12}$eV m according to the spin-splitting energy. One great benefit from using 2DEG lies in that the RSOC can be nicely tuned by electric fields as predicted by Shanavas et al. \cite{shanavas}, and demonstrated by experiments\cite{nakamura,caiglia}. The asymmetric 2D electron system can be realized in ultra-thin metallic layers and surfaces. The current-induced spin polarization on free metal surfaces has also been systematically studied on the surfaces of Au, Cu, Pt, Pd, Ta and W by spin-polarization positron beam\cite{zhang}. Besides these surfaces, metallic interfaces can also possess this asymmetry as shown in Pt/Co(0.6 nm)/AlO$_x$\cite{miron} at room temperature with $\alpha_R=10^{-10}$eV m.  
\par Following the Rashba Hamiltonian, an associated anomalous charge current $ \vec{j}_c=-e{[\vec{r}, H_{RSOC}]}/{i\hbar}=-e\alpha_R(\vec{z}\times\vec{\sigma})/\hbar$ arises from the non-equilibrium spin density $\vec{\sigma}$. This spin galvanic effect is also called the Inverse Rashba Edelstein effect (IREE)\cite{edelstein}. Recently, experimental evidences for a large spin-charge conversion at the Bi/Ag\cite{fert13,sangiao}, Ag/Sb\cite{zhang15} and LaAlO$_3$/SrTiO$_3$\cite{lesne} interfaces were found by the use of microwave spin pumping to inject spin currents from the NiFe layer into the heterointerface. 

\par The 2DEG can be also formed at the surfaces of transitional metal oxides\cite{meevasana,chang,wang}, which exhibits extraordinary properties such as tunable insulator-metal transitions\cite{liu}, large anisotropic magnetoresistance\cite{bruno}, etc.. Due to the asymmetric geometry, the Rashba SO coupling is anticipated. Evidence of the Rashba spin splitting in the quasi 2DEG formed at the (001) surface of SrTiO$_3$ (STO) single crystal was found from weak localization or antilocalization analysis of the low-temperature magnetoresistance\cite{nakamura}. The spin splitting energy is 0.1-0.3 meV. Later on, a giant spin splitting of about 100 meV at the Fermi level in the two-dimensional electron gas formed at the surface of STO was observed at 20 K, which goes beyond the expectation of the normal Rashba effect: Considering the band bending $\sim 300$ meV amounts to an electric field $\sim100$MV m$^{-1}$, which will give a spin splitting of only about 10$^{-5}$ meV. The surface corrugations may result in a greatly enhanced Rashba parameter that increases the spin splitting \cite{radovic}. As experiments clearly show the presence of SO coupling and spin splitting in the 2DEG on STO surfaces, the IREE can also be anticipated. This is the main result of this work, where a clear evidence of the voltages induced by the non-equilibrium spins pumped from NiFe films into the conducting STO surfaces after Ar$^+$ irradiation was observed at room temperature. It shows that the spin-to-charge conversion can be comparable with that of the Bi/Ag bilayer. The results support a large Rashba coupling constant in this 2DEG, which holds great potential to explore spintronic physics in such treated oxide surfaces.          
 \par The substrate used in this study is the polished single-crystal STO (001) with dimensions of 10 mm $\times$ 5 mm $\times$ 0.2 mm, which is mounted in a vacuum chamber and bombarded with Ar$^+$ generated by a home made Hall ion source. The implantation dose is about 1.2$\times$10$^{22}$ cm$^{-2}$. Subsequently, a 20-nm-thick ferromagnetic NiFe layer was magnetron sputtered on the non-treated and ion-irradiated STO substrate with the argon pressure at 0.5 Pa and the RF power at 100 W. 
 \par The microwave measurement was done by our shorted microstrip fixture which can work up to 8 GHz at the room temperature. In order to put the samples at the same positions in the fixture and minimize the differences of the experienced microwave field before and after sample flipping, we covered the samples with virgin STO of  the same dimensions as the substrate. We obtained the voltage by lock-in techniques (SR830, Standford Research System) with microwave source power provided by Rohde \& Schwarz (SMB 100A). At the fixed microwave frequency, we sweep the static magnetic field so that FMR was observed. 
 \par To determine the thickness of the conductive layer, a wedge cross-section etched from a photoresist patterned sample by a Ar$^+$ ion beam of 5 keV for 30 seconds (PIPS,  Gatan Inc.).  The tilt angle between the Ar$^+$ ion beam and the sample surface was set to be as low as possible, around 6 degrees. Current AFM was measured at room temperature by an atomic force microscope (AFM I, Attocube) equipped with a low noise current amplifier (SR570, Stanford Research System). During the CAFM measurement, a dc bias of 2 V was applied to the sample, while the conductive diamond coated probe (CDT-CONTR, Nanosenors) was virtually grounded. 
 \par The temperature dependent  transport properties were measured in the Hall bar configuration. The current source is provided by Keithley 6221A and the voltage is measured by Keithley 2182A, while the temperature is controlled by Lakeshore 335.   
 

\par It is well demonstrated that the surface of insulating STO transforms into metallic after proper Ar$^+$ irradiation. It is due to the formation of oxygen vacancies after that \cite{meevasana, bruno}.  At the same time, the transparency is also decreased. 
\par The conducting layer  is estimated from the conductive AFM (CAFM) measurement of the wedge cross-section of STO as schematically shown in Fig.\ref{fig:con}. (a). From the simultaneously obtained topography and current images and the averaged profiles (Fig.\ref{fig:con} (b)-(e)), it is found that the thickness of conductive layer is within 5 nm below the STO surface. As the trace distance of CAFM probe on the wedge cross-section was much larger than the thickness of the conductive layer, the limit of the spatial current resolution of CAFM, typically around 5 nm, was overcame. Our results suggest that a conducting layer is formed on the STO surface by Ar$^+$ bombardment.
The conductivity is exponentially dependent on the depth. It is in agreement of depth dependence of oxygen concentration after irradiation.\cite{wang}    
\begin{figure}
	\caption{ (a) Schematics of the CAFM measurement on the wedge cross-section of STO. The topography (b) and current (c) images was obtained simultaneously, the scan size is 2$\times$0.5 $\mu$m$^2$. The profiles of topography (d) and current (e) along the wedge cross-section was averaged from 100 scan lines.}\label{fig:con}
\end{figure}

\begin{figure}
 	\includegraphics[width=0.48\textwidth]{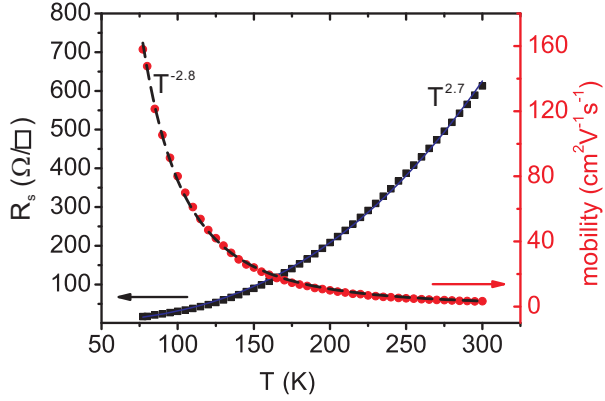}
 	\caption{The temperature dependence of the sheet resistivity R$_s$ and the electron Hall mobility of Ar$^+$ irradiated SrTiO$_3$ (001) surfaces. The fitting curves of the exponential function of temperature (T) were also shown by the lines on the data points. }\label{fig:r-t}
 \end{figure}

As shown in Fig. \ref{fig:r-t}, the sheet resistance  increases as the temperature increases, which is evident of metallic behavior. The data can be fitted to $T^n$ with $n\sim2.7$.  The sheet resistance (R$_s$) reaches $\sim$600 $\Omega/\Box$ at room temperature. The electron Hall mobility reaches $\sim 160$ cm$^2$V$^{-1}$s$^{-1}$  at 77 K. Up to the room temperature, the temperature dependent electron mobility can be best fitted to $ T^m$ with $m\sim -2.8$. The behavior deviates from the phonon dominant scattering of non-degenerated 2DEG, which gives $m=-1\sim-1.5$\cite{stern}.  Yet, the origin of the surface 2DEGs on STO remains unclear, or even controversial\cite{meevasana,plumb,sivadas}. 
  
\par
\begin{figure}
 	\includegraphics[width=0.5\textwidth]{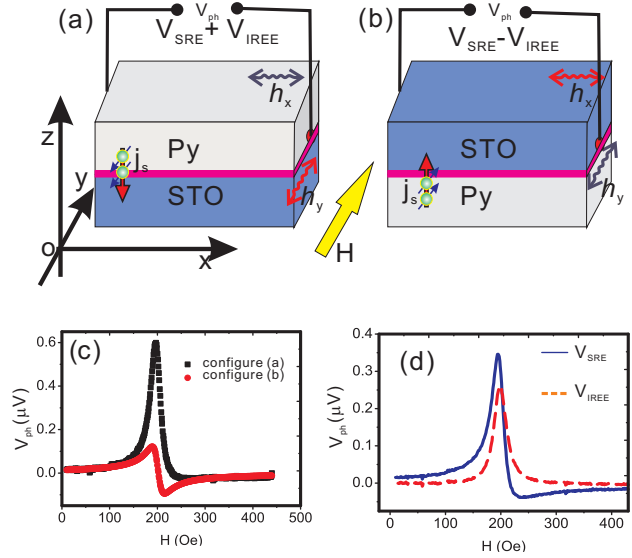}
 	\caption{Measurement of the IREE voltages. The measurements were done in two steps as schematically  shown in (a) and (b). The corresponding measured photon voltages before (up) and after (dn) sample flipping as the function of the magnetic field $H$ are shown in (c) . The voltages of SRE and IREE obtained from the curves (c) are shown in (d) . }\label{fig:vishe}
 \end{figure}

The spin transport properties were determined by the flip-chip method using a shorted micro-strip fixture which was recently proposed by us \cite{wxz16}   as shown schematically in Fig.\ref{fig:vishe}(a) and (b). The photon voltage is usually a superposition of the spin rectification effect(SRE) in NiFe, which comes from the rectification of the dynamic current and the oscillating magneto-resistance driven by a dynamic magnetic field, and the IREE in the metallic surface of STO. According to Rashba spin orbital Hamiltonian the spin current can be written as $ \vec{j}_c=-e\alpha_R(\vec{z}\times\vec{\sigma})/\hbar$. We can see that $\vec{j}_c$ is an odd function of the surface normal $\vec{z}$, while the spin rectification effect (SRE) is independent of it. Thus, IREE and SRE may be separated by comparing their combined voltages after inverting the samples in the z plane. Specifically, the voltage due to the IREE is $V_{IREE}=(V_{up}-V_{dn})/2$ and  $V_{SRE}=({V_{up}+V_{dn}})/{2}$, where the $V_{up,dn}$ are the photon voltages measured before and after the sample is flipped as shown in Fig.\ref{fig:vishe}(a) and (b). The voltage measured at 4.2 GHz  is shown in Fig.\ref{fig:vishe}(c). The respective contributions of SRE and IREE voltages are quite obvious in the figure as indicated by the difference of the two curves before and after chip flipping. As can be seen from the figure, the SRE voltage is a combination of the static magnetic field $H$-dependent symmetric and antisymmetric Lorentzian curves $V_{SRE}=V_L\cdot L(H)+V_D\cdot D(H)$, with
\begin{eqnarray}
D(H)=\frac{2\Delta H(H-H_r)}{4(H-H_r)^2+\Delta H^2} \\
L(H)=\frac{\Delta H^2}{4(H-H_r)^2+\Delta H^2},
\end{eqnarray}
where $H_r$ and $\Delta H$ are the ferromagnetic resonant field and linewidth at ferromagnetic resonance (FMR), respectively. Clearly seen from the curve in Fig.\ref{fig:vishe} (d), there is symmetric Lorentzian component from IREE. The IREE voltage is of the same order of magnitude as SRE.
 \par
\begin{figure}
 	\includegraphics[width=0.5\textwidth]{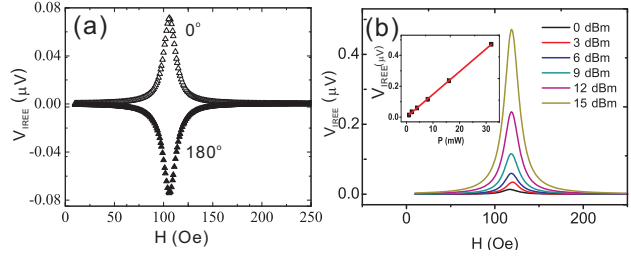}
 	\caption{The IREE voltages when the field is reversed from $y$ axis (0$^\circ$) to $-y$ axis (180$^\circ$) (a) and the power dependent IREE voltage at 4.2 GHz (b).}\label{fig:v-p}
 \end{figure}

 We also measured the voltages when the magnetic field is  reversed from $y$ (0$^\circ$) axis to $ -y$ axis (180$^\circ$). The IREE voltages are shown in Fig.\ref{fig:v-p}(a) as a function of the static magnetic field. The reversal of the magnetic field causes the reversal of the spin polarization $\vec{\sigma}$, thus changes the voltage to switch signs from positive to negative, in agreement with the theoretical prediction of spin current conversion to the charge current.   
 \par The power dependent voltage of IREE is shown in Fig.\ref{fig:v-p}(b). Because the V$_{IREE}$ is proportional to the spin current $j_s$, which is in turn proportional to the square of the microwave magnetic field $h_{eff}^2\propto P$, this leads to the linear dependence of the $V_{IREE}$ on the microwave power. This linear dependence of the photon voltage and microwave power is shown in the inset of Fig.\ref{fig:v-p}(b). 
\par The trajectories of the resonant peaks, i.e. the resonant frequency ($f_r$) vs. the resonant field ($H_r$), from symmetric Lorentzian lines of IREE and SRE are basically the same, which reflect the FMR condition in Py films. They are fitted to the Kittel's formula 
 \begin{equation}
 f_r=\frac{\gamma}{2\pi}\sqrt{(H_r+H_k)(H_r+H_k+4\pi M_s)},
 \end{equation}
 where $\gamma$ is the gyromagnetic ratio. With the effective saturation magnetization $4\pi M_s\sim 11$ kOe and magnetocrystalline anisotropic constant $H_k\sim 0.84$ Oe, the data can be nicely fitted as shown by the red line in the figure. The amplitude $V_L$ and $V_D$ of SRE changes with the frequency, because of the variation of the phase between the dynamic current $j$ and magnetic field\cite{hu11}. This frequency dependent amplitude mixing of $V_L$ and $V_D$ prevents us from simply taking the symmetric component of the voltage as contribution from IREE, while the asymmetric one from the SRE.
 \par
\begin{figure}
  	\includegraphics[width=0.4\textwidth]{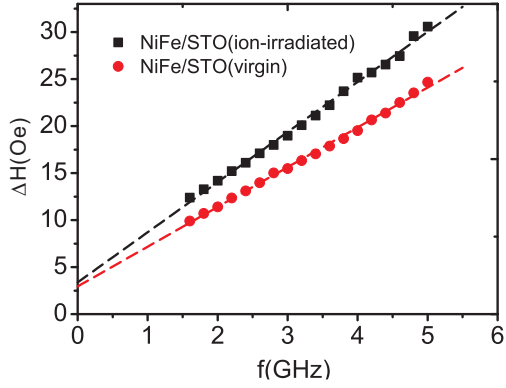}
  	\caption{The frequency dependent FMR line width ($\Delta H$) of NiFe films on the virgin and irradiated STO. }\label{fig:deltaHf}
  \end{figure}

 The IREE is the process of conversion of spin current $j_s$ to the charge current $j_c$ via RSOC. Its efficiency can be measured. The coefficient $\lambda_{IREE}$ connects the two quantity\cite{fert13}:
 \begin{equation}
 j_c=\lambda_{IREE}j_s
 \end{equation}     
 The spin current $j_s$ pumped from NiFe at FMR can be obtained by measuring the spin mixing conductance $G_{mix}$, which is related to the increase of the NiFe damping due to spin pumping. The linewidth ($\Delta H$) from the virgin and irradiated STO substrate are shown in Fig. \ref{fig:deltaHf}, which clearly show an increase when spin pumping is present in the conducting STO/NiFe bilayer. They can be linearly fitted by $\Delta H=\Delta H_0+4\pi  \alpha f/\gamma$. The two lines give $\alpha_{0}=  0.01025$  and $\alpha_{eff}=0.01292$ for the virgin and irradiated substrate, respectively. The zero-freqency linewidth $\Delta H_0$ comes from inhomogeneous broadening due to imperfections in the films \cite{heirich}. Extrapolation of both lines to $f=0$ GHz gives $\Delta H_0\sim 3$ Oe. The similar $\Delta H_0$ obtained for both samples indicates that the films grown on both substrates are of the same crystalline quality. The increased line width for the irradiated STO/NiFe sample, which is proportional to the frequency comes from the spin pumping, which is characterized by the spin mixing conductivity,   
  \begin{equation}
  G_{mix}=\frac{4\pi M_{s}t_F}{g\mu_B}(\alpha_{eff}-\alpha_0),
  \end{equation}
  where $M_s$, $t_F$ and $\mu_B$ are the effective saturation magnetization, the thickness of the ferromagnetic layer, and the Bohr magneton, respectively. The damping constants $\alpha_{eff(,0)}$ are the values taken from Py on irradiated and virgin STO, respectively. With the data obtained from the above, the spin mixing conductivity is $2.98\times10^{19}$ m$^{-2}$, which is comparable with that of Py/GaAs\cite{ando}, e.g. $2.31\times10^{19}$m$^{-2}$.  

 \par The spin current $j_s$ is calculated as
 \begin{equation}
 j_s=\frac{G_{mix}\gamma^2\hbar h^2_{rf}}{8\pi\alpha_{eff}^2}[\frac{4\pi M_s\gamma+\sqrt{(4\pi M_s\gamma)^2+4\omega^2}}{(4\pi M_s\gamma)^2+4\omega^2}]\frac{2e}{\hbar}.
 \end{equation}
 The calculated value of $j_s$ is 10.8 kA m$^{-2}$ in our sample at $f=4.8$ GHz, $\omega=2\pi f$ and $h_{rf}= 0.13$ Oe. The spin current is absorbed in  the conduction surface of STO and produces nonequilibrium spin density. By the IREE, it is converted to the charge current density $j_c$ flowing through the 2D conducting layer, which can be estimated by 
\begin{equation}
j_c=\frac{V_{IREE}}{R_sl},
\end{equation} 
 where $R_s$ and $l$ are the sheet resistance of the sample and distance between the electrodes, respectively. Taking the experimental values of these parameters, we obtained $\lambda_{IREE}\sim 0.23$ nm. This length is in the same order as $\lambda_{IREE}=0.33$ nm in Ag/Bi interfaces reported by S$\acute{a}$nchez et al. \cite{fert13} .
 \par As reported by Santander-Syro et al\cite{radovic}, the spin splitting of 2DEG in STO is very large, reaching more than 100 meV. This gives a Rashba coefficient of $\alpha_R\sim 10^{-12}$ eV m. As in the limit of strong spin-moment coupling, the $\lambda_{IREE}\sim \alpha_R\tau_s/\hbar$, where $\tau_s$ is the effective relaxation time describing the ratio between spin injection and spin-moment scattering. Taking our $\lambda_{IREE}= 0.23$ nm, we get $\tau_s=0.15$ ps, which is reasonable when considering the rough surface of STO after Ar$^+$ irradiation. In terms of level broadening, one obtains $\hbar/\tau_s=4.4$ meV, which is only a percentage of spin splitting on the corrugated STO surfaces\cite{radovic}. The results obtained so far supported the huge Rashba splitting picture of STO surfaces, where multiorbital, especially the $d$-orbital  Rashba physics and surface atomic configuration are both critical \cite{nakamura}.  
\par  In summary, we have measured the electric voltage on the metallic conducting (001) surface of SrTiO$_3$, converted from the spin current which is pumped from the NiFe layer by microwave excitation at ferromagnetic resonance. The conversion efficiency is on the same order as the interfaces in metallic multilayers. The spin-orbital coupling constant obtained from our measurements support the giant spin splitting picture from ARPES measurements.  This extends the range of potential materials for spin-current detector without magnetic materials. In addition, the 2DEG allows the spin orbital coupling to be presented in a controllable way.  
\par Financial support from NSFC(61471095, V01435208), ``863''-projects (2015AA03130102) and Research Grant of Chinese Central Universities (ZYGX2013Z001) are acknowledged.

\end{document}